\documentclass[12pt]{article}
\usepackage{epsf}
\usepackage{graphics}

\begin{document}
\title
{Another possible interpretation of SN 1a data - without
kinematics: \\ Comments on the paper astro-ph/0402512 by A. Riess
et al.}
\author
{Michael A. Ivanov \\
Physics Dept.,\\
Belarus State University of Informatics and Radioelectronics, \\
6 P. Brovka Street,  BY 220027, Minsk, Republic of Belarus.\\
E-mail: ivanovma@gw.bsuir.unibel.by.}
\date{March 3, 2004}
\maketitle

\begin{abstract}
It is shown here that for redshifts $z < 0.5$ the luminosity
distance, which is predicted in author's model astro-ph/0005084
v2, fits well supernova observational data of astro-ph/0402512 by
A.Riess et al. Discrepancies for higher $z$ would be explained in 
the model as a result of specific deformation of SN spectra due to a discrete
character of photon energy losses. The model does not
require any dark energy; it is based on the conjecture that
gravitons are super-strong interacting particles fulfilling a flat
non-expanding universe.
\end{abstract}

In a presence of the graviton background with the Planckian
spectrum ($T$ is an effective temperature of the background),
which is considered in a flat space-time, an energy of any photon
should decrease with a distance $r,$ and we have for a redshift
$z$ \cite{1}: $z=\exp(ar)-1.$ Here $a=H/c$ with the Hubble
constant: $H= (1/2\pi) D \cdot \bar \epsilon \cdot (\sigma
T^{4}),$ where $\bar \epsilon$ is an average graviton energy,
$\sigma$ is the Stephan-Boltzmann constant, and $D$ is some new
dimensional constant. It is necessary to accept for a value of
this constant: $D \sim 10^{-27} m^{2}/eV^{2}.$ In this approach,
the Newton constant $G$ is connected with $H$ \cite{2}, and one
can compute: $H= 3.026 \cdot 10^{-18}s^{-1}=94.576 \ km \cdot
s^{-1} \cdot Mpc^{-1}$ by $T=2.7 K.$
\par
An additional relaxation of any photonic flux due to non-forehead
collisions of gravitons with photons leads to the luminosity
distance $D_{L}:$
\begin{equation}
D_{L}=a^{-1} \ln(1+z)\cdot (1+z)^{(1+b)/2} \equiv a^{-1}f_{1}(z),
\end{equation}
where $b= 3/2+2/\pi =2.137$ is a computable constant of this
model.
\par
\begin{figure}[th]
\epsfxsize=12.98cm \centerline{\epsfbox{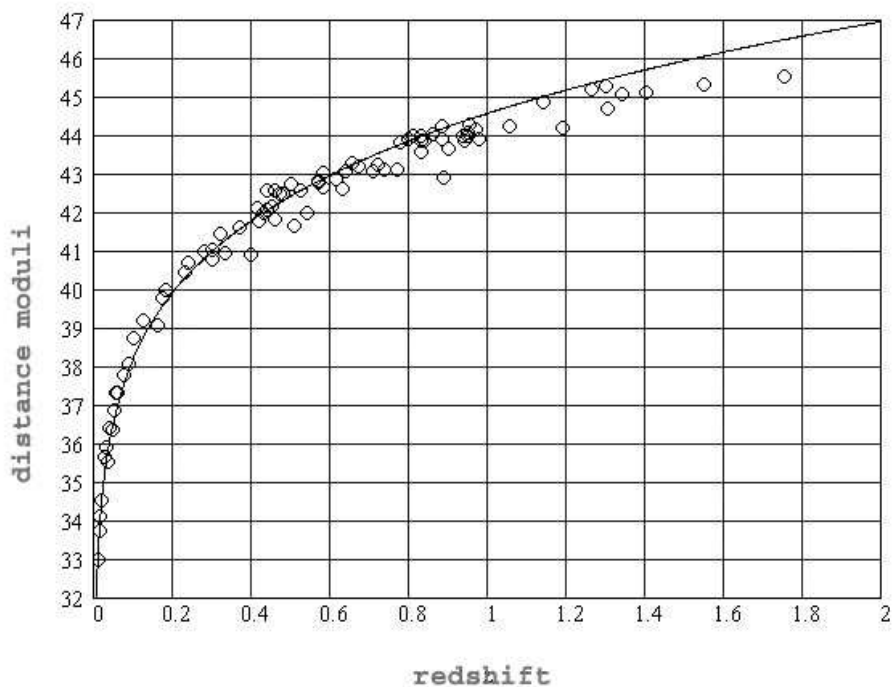}}
\caption{Comparison of the theoretical values of distance moduli
$\mu_{0}(z)$ (solid line) with observations (points) from \cite{3}
by Riess et al.}
\end{figure}
To compare a form of this dependence by unknown, but constant $H$,
with supernova data by Riess et al. \cite{3}, one can introduce
distance moduli $\mu_{0} = 5 \log D_{L} + 25 = 5 \log f_{1} +
c_{1}$, where $c_{1}$ is an unknown constant (it is a single free
parameter to fit the data); $f_{1}$ is the luminosity distance in
units of $c/H$. In Figure 1, the Hubble diagram $\mu_{0}(z)$ is
shown with $c_{1}=43$ to fit observations for low redshifts;
observational data (82 points) are taken from Table 5 of \cite{3}.
The predictions fit observations very well for roughly $z < 0.5$.
\par
Discrepancies between predicted and observed values of
$\mu_{0}(z)$ are obvious for higher $z$: we see that observations
show brighter SNe that the theory allows, and a difference
increases with $z$. It is better seen on Figure 2 with a linear
scale for $f_{1}$; observations are transformed as $\mu_{0}
\rightarrow 10^{(\mu_{0}-c_{1})/5}$ with the same
$c_{1}=43$.\footnote{A spread of observations raises with $z$; it
might be partially caused by quickly raising contribution of a
dispersion of measured flux: it should be proportional to
$f_{1}^{6}(z)$.}
\begin{figure}[th]
\epsfxsize=12.98cm \centerline{\epsfbox{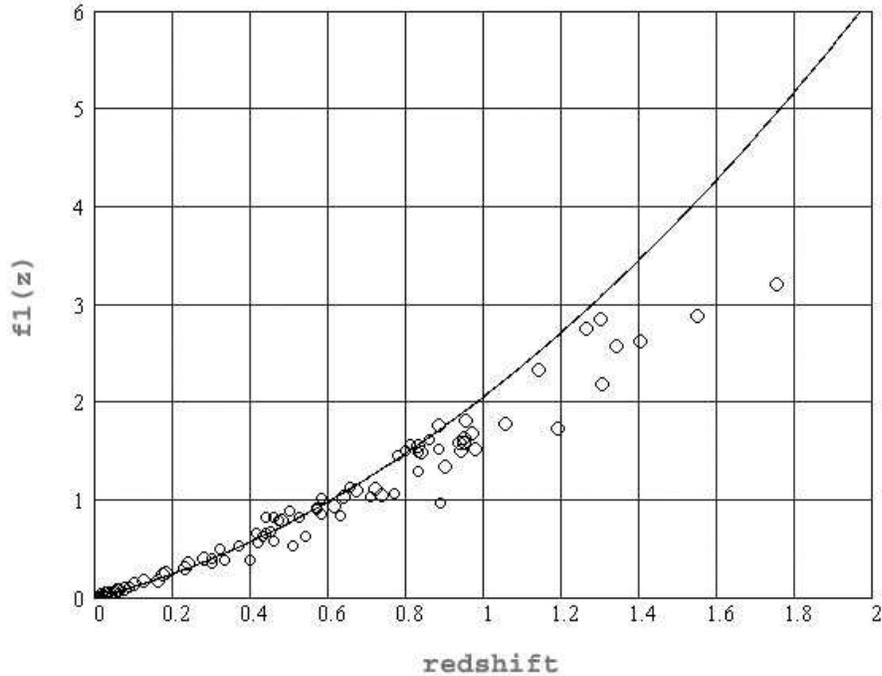}}
\caption{Predicted values of $f_{1}(z)$ (solid line) and
observations (points) from \cite{3} transformed to a linear scale}
\end{figure}
\par
It would be explained in the model as a result of specific
deformation of SN spectra due to a discrete character of photon
energy losses. Today, a theory of this effect does not exist, and
I explain its origin only qualitatively \cite{4}. For very small
redshifts $z,$ only a small part of photons transmits its energy
to the background. Therefore any red-shifted narrow spectral strip
will be a superposition of two strips. One of them has a form
which is identical with an initial one, its space is proportional
to $1-n(r)$ where $n(r)$ is an average number of interactions of a
single photon with the background, and its center's shift is
negligible (for a narrow strip). Another part is expand, its space
is proportional to $n(r),$ and its center's shift is equal to
$\bar \epsilon _{g} /h$ where $\bar \epsilon _{g} $   is an
average energy loss in one act of interaction. An amplitude of the
red-shifted step should linear raise with a redshift. For big $z,$
spectra of remote objects of the Universe would be deformed. A
deformation would appear because of multifold interactions of a
initially-red-shifted part of photons with the graviton
background. It means that the observed flux within a given
passband would depend on a form of spectrum: the flux may be
larger than an expected one without this effect if an initial flux
within a next-blue neighbour band is big enough - due to a
superposition of red-shifted parts of spectrum. Some other
evidences of this effect would be an apparent variance of the fine
structure constant \cite{5} or of the CMB temperature \cite{6}
with epochs. In both cases, a ratio of red-shifted spectral line's
intensities may be sensitive to the effect.
\par
In conclusion, it may be noted that grand kinematic maneuvers with
the entire universe (acceleration after deceleration etc) need an
engine to do them (given an engine, one may dream about a control
system to operate it). A sea of super-strong interacting gravitons
would appear to be a reasonable alternative; perhaps, it does not
need any kinematics. New gravitational physics may underlie the
remarkable results of astrophysical observations by A. Riess et
al.

\end{document}